# Reduced-Scaling Double Hybrid Density Functional Theory with Rapid Basis Set Convergence through Localized Pair Natural Orbital F12

Nisha Mehta and Jan M. L. Martin*

**ABSTRACT:** Following earlier work [Mehta, N.; Martin, J. M. L. *J. Chem. Theory Comput.* **2022**, 10.1021/acs.jctc.2c00426] that showed how the slow basis set convergence of the double hybrid density functional theory can be obviated by the use of F12 explicit correlation in the GLPT2 step (second order Görling-Levy perturbation theory), we demonstrate here for the very large and chemically diverse GMTKN55 benchmark suite that the CPU time scaling of this step can be reduced (asymptotically linearized) using the localized pair natural orbital (PNO-L) approximation at negligible cost in accuracy.

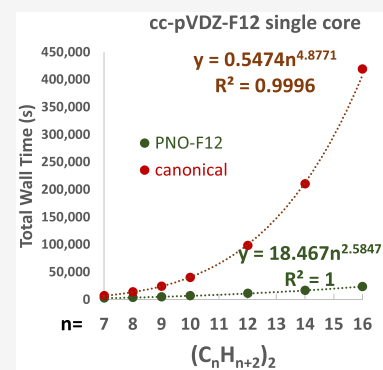

Despite the enormous success of density functional theory (DFT),[1,2] the exact exchange-correlation functional continues to be elusive, and accurate density functional approximations (DFAs) are highly desirable in computational chemistry. On Perdew's "Jacob's Ladder" of DFT,[3] the fifth rung corresponds to introducing dependence on unoccupied (virtual) orbitals; one special case of fifth-rung functionals are double hybrid DFAs.[4−13] These combine semilocal exchange and correlation from DFT with nonlocal Fock exchange and GLPT2 (second-order Görling-Levy perturbation theory[14]) nonlocal correlation contributions. Indeed, double hybrid density functionals (DHDFs) are known to be the most accurate DFAs, approaching composite wave function theory schemes such as G3 and G4 theories.[15−17]

Although the basis set convergence of double hybrids is faster than that of wave function *ab initio* (WFT) methods, they inherit the slow basis set convergence of MP2 ($\propto L^{-3}$ with $L$ being the highest angular momentum in the basis set) to some degree.

WFT basis set convergence can be greatly accelerated using explicitly correlated methods[18−20] in which some geminal terms that explicitly depend on interelectronic distances are added to the one-particle basis set. For instance, Kutzelnigg and Morgan[21] showed that, for a two-electron system, the singlet-coupled pair correlation energies converge as $\propto L^{-7}$ for explicitly correlated methods, compared to $\propto L^{-3}$ for orbital-only calculations.

While early explicitly correlated WFT studies[22,23] employed simple $r_{12}$ geminals, the F12 geminal,[24] $(1 − \exp \gamma r_{12})/\gamma$, has become the *de facto* standard for explicitly correlated WFT. The introduction of density fitting and auxiliary basis sets sped up the integral evaluation to the point that F12 calculations became practically feasible.[25−27]

Practical experience with MP2-F12 and various approximations[28−31] to CCSD(T)-F12 has shown that, using basis sets specifically developed for F12 calculations,[32−34] the basis set convergence is drastically faster than for conventional orbital calculations. Thus, F12 approaches have increasingly become a mainstay of high-accuracy WFT; see, for instance, refs 35−45.

We have recently shown,[46] for the large and chemically diverse GMTKN55 benchmark suite,[6] that using MP2-F12 in a basis of Kohn−Sham orbitals accelerates basis set convergence of double hybrid DFT to the point that even *spd* basis sets like cc-pVDZ-F12[32] are quite close to the basis set limit, and the *spdf* cc-pVTZ-F12[32] effectively reaches it.

Nevertheless, two problems remain. First of all, CPU times for the MP2-F12 step, which scale[47] as the fifth power of system size (sixth power if F12 amplitudes are optimized in an orbital invariant manner) will become prohibitive for medium−large systems. Second, in some implementations (such as that in MOLPRO[48]), F12 methods require a very large amount of scratch storage space, the reading and writing of which creates an I/O bottleneck.

Table 1. Statistical Analysis (kcal/mol) of the Basis Set Convergence in Localized B2GP-PLYP-F12-D3(BJ) Calculations for the GMTKN55 Database and Its Categories

|  | WTMAD2 | THERMO | BARRIERS | LARGE | CONF | INTERMOL |
|---|---|---|---|---|---|---|
| | | Relative to the Ref 6 Reference Data | | | | |
| VDZ-F12 | 3.080 | 0.590 | 0.327 | 0.662 | 0.636 | 0.864 |
| VDZ-F12*[a] | 3.073 | 0.583 | 0.325 | 0.662 | 0.636 | 0.867 |
| VTZ-F12 | 3.030 | 0.590 | 0.324 | 0.659 | 0.599 | 0.858 |
| VTZ-F12*[a] | 2.961 | 0.587 | 0.322 | 0.659 | 0.599 | 0.794 |
| V{D,T}Z-F12 | 3.025 | 0.588 | 0.327 | 0.655 | 0.593 | 0.863 |
| V{D,T}Z-F12*[a] | 2.951 | 0.586 | 0.324 | 0.655 | 0.593 | 0.793 |
| | Relative to the B2GP-PLYP-F12-D3(BJ)/V{T,Q}Z-F12* Basis Set Limits from ref 46 | | | | | |
| VDZ-F12 | 0.642 | 0.086 | 0.052 | 0.093 | 0.135 | 0.277 |
| VDZ-F12*[a] | 0.633 | 0.074 | 0.048 | 0.093 | 0.135 | 0.284 |
| VTZ-F12 | 0.322 | 0.032 | 0.019 | 0.033 | 0.057 | 0.182 |
| VTZ-F12*[a] | 0.233 | 0.023 | 0.017 | 0.033 | 0.057 | 0.103 |
| V{D,T}Z-F12 | 0.300 | 0.033 | 0.023 | 0.033 | 0.048 | 0.163 |
| V{D,T}Z-F12*[a] | 0.262 | 0.026 | 0.021 | 0.033 | 0.048 | 0.134 |
| | Relative to Canonical B2GP-PLYP-F12-D3(BJ) in the Same Basis Set from ref 46 | | | | | |
| VDZ-F12 | 0.513 | 0.046 | 0.034 | 0.062 | 0.113 | 0.258 |
| VDZ-F12*[a] | 0.525 | 0.045 | 0.032 | 0.062 | 0.113 | 0.273 |
| VTZ-F12 | 0.333 | 0.024 | 0.017 | 0.038 | 0.066 | 0.188 |
| VTZ-F12*[a] | 0.269 | 0.024 | 0.018 | 0.038 | 0.066 | 0.123 |
| V{D,T}Z-F12 | 0.335 | 0.023 | 0.018 | 0.041 | 0.073 | 0.180 |
| V{D,T}Z-F12*[a] | 0.317 | 0.023 | 0.020 | 0.041 | 0.073 | 0.161 |

[a]The VnZ-F12 (where n = D,T) basis set was used for Ne-containing systems in RG18 due to numerical problems. Only for canonical VTZ-F12, computational resource limitations forced substitution of VDZ-F12 for the UPU23 subset.

Localized pair natural orbital (PNO-L) WFT approaches, such as DLPNO-CCSD(T) of Neese and co-workers,[49] PNO-CCSD(T) of Ma and Werner,[50] and LNO-CCSD(T) of Nagy and Kállay,[51] are gaining increasing acceptance, as their size scaling is asymptotically linear; the same can be achieved for MP2 (and in fact this is an intermediate step in the aforementioned calculations). Moreover, when used in tandem with F12 approaches, they combine rapid basis set convergence with gentle system-size scaling, such as in the PNO-MP2-F12 and PNO-CCSD(T)-F12 approaches of Ma and Werner.[50,52]

It stands to reason that PNO-LMP2-F12 in a basis of Kohn−Sham orbitals might be a robust and computationally efficient way around the scaling and storage bottlenecks of double hybrid DFT (i.e., PNO-DHDF-F12). Assessing the performance of PNO-DHDF-F12 against canonical benchmark results is essential for confirming its robustness. We will show below that, when applied to GMTKN55, PNO-DHDF-F12 provides essentially similar accuracy to canonical DHDF-F12, but at a much reduced computational cost for large molecules. The latter fact greatly increases the scope where PNO-DHDF-F12 can be applied.

This study focuses on the GMTKN55 database for general main-group thermochemistry, kinetics, and noncovalent interactions. It consists of 55 problem sets comprising 1505 relative energies, which entail 2462 unique single-point energy calculations. These 55 sets can be grouped into five top-level categories: basic properties and reactions of small systems ("Thermo"), reaction energies of large systems and isomerizations ("Large"), barrier heights ("Barrier"), intermolecular noncovalent interactions ("Intermol"), and intramolecular noncovalent interactions ("Conf"). For more details, see ref 6 and the references therein.

The performance metric used here, as in ref 6 and subsequent studies, e.g., refs 7−10, is the weighted total mean absolute deviation (WTMAD2) as defined in ref 6:

$$\text{WTMAD2} = \frac{1}{\sum_{i=1}^{55} N_i} \sum_{i=1}^{55} N_i \frac{56.85 \text{ kcal/mol}}{|\overline{\Delta E_i}|} \text{MAD}_i \quad (1)$$

where $N_i$ denotes the number of systems in each test set, $|\overline{\Delta E_i}|$ is the mean absolute value of all reference energies from $i = 1$ to 55, and $\text{MAD}_i$ is the mean absolute deviation of the calculated and reference energies. We also consider the decomposition of WTMAD2 into GMTKN55's five top-level subcategories.

All electronic structure calculations were performed using the MOLPRO2022[48] package on the Faculty of Chemistry's HPC cluster "ChemFarm" at the Weizmann Institute of Science. The B2GP-PLYP-D3(BJ) simple double hybrid[53,54] was considered as a "proof of principle". Computational details in this work largely follow those in our previous study (ref 46). All of the KS and PNO-LMP2-F12 steps were performed using density fitting (DF), and the default PNO settings were applied throughout. We considered here the cc-pVnZ-F12 (VnZ-F12 in short) basis sets,[32] augmented versions thereof,[33] and cc-pVnZ-PP-F12[34] basis sets for the heavy p-block, where n = D,T. Throughout the manuscript PNO-DHDF-F12 refers to DHDF-F12 calculations with the DF-PNO-LMP2-F12. For the CABS (complementary auxiliary basis set),[55] we used the cc-pVnZ-F12/OptRI auxiliary basis sets;[56] for Coulomb-exchange ("JK") fitting, those of Weigend;[57] finally, for RI-MP2, the MP2FIT sets of Hättig and co-workers.[58,59] The self-consistent-field (SCF) calculations were carried out with a convergence criterion of $10^{-9} E_h$. All SCF calculations were conducted with MOLPRO's default integration grid combinations but with `gridthr` tightened to $10^{-9}$. The fixed-amplitude "3C(FIX)" approximation[24,28] was employed throughout.

Table 2. Relative Wall-Clock Times for the PNO-B2GP-PLYP-F12, B2GP-PLYP-F12, and B2GP-PLYP Calculations for $(C_nH_{n+2})_2$[a]

| | B2GP-PLYP-F12 | | B2GP-PLYP | | | | | | B2GP-PLYP-F12 | |
|---|---|---|---|---|---|---|---|---|---|---|
| | PNO-F12 | canonical | | | | | | | PNO-F12 | canonical |
| | VDZ-F12 | VDZ-F12 | {T,Q}ZVPP | {T,Q}ZVPPD | V{T,Q}Z | haV{T,Q}Z | haV{Q,5}Z | AV{T,Q}Z | VTZ-F12 | VTZ-F12 |
| (butane)$_2$ | 1.00 | 1.20 | 1.10 | 1.77 | 1.90 | 2.22 | 8.61 | 4.18 | 3.50 | 4.22 |
| (pentane)$_2$ | 1.00 | 1.58 | 1.10 | 1.82 | 1.43 | 2.34 | 9.18 | 4.38 | 3.52 | 5.79 |
| (hexane)$_2$ | 1.00 | 2.05 | 1.18 | 1.96 | 1.50 | 2.53 | 10.13 | 4.72 | 3.57 | 7.94 |
| (heptane)$_2$ | 1.00 | 2.61 | 1.24 | 2.12 | 1.60 | 2.74 | 11.06 | 5.12 | 3.62 | 10.26 |
| (n-octane)$_2$ | 1.00 | 3.52 | 1.35 | 2.33 | 1.72 | 2.99 | 12.08 | 5.56 | 3.63 | 13.57 |
| (n-nonane)$_2$ | 1.00 | 4.52 | 1.46 | 2.53 | 1.87 | 3.27 | 13.36 | 6.06 | 3.63 | 17.40 |
| (n-decane)$_2$ | 1.00 | 5.70 | 1.61 | 2.81 | 2.04 | 3.60 | 14.81 | 6.61 | 3.64 | 21.87 |
| (n-dodecane)$_2$ | 1.00 | 8.71 | 1.95 | 3.42 | 2.46 | 4.31 | 18.22 | 7.98 | 3.67 | |
| (n-tetradecane)$_2$ | 1.00 | 12.39 | 2.49 | 4.11 | 2.95 | 5.94 | 23.59 | 11.27 | 3.60 | |
| (n-hexadecane)$_2$ | 1.00 | 17.60 | 2.85 | 5.02 | 3.48 | 6.62 | 27.35 | 11.31 | 3.64 | |

[a]All timings on a single Intel Haswell 2.4 GHz core in 256 GB RAM and with a 3.6TB striped solid state scratch disk. Timing is shown relative to PNO-B2GP-PLYP-F12/VDZ-F12. Timings for extrapolated B2GP-PLYP/{T,Q}ZVPP and the like correspond to the sum of TZVPP and QZVPP and so forth.

In both ref [46] and the present work, one subset, C60ISO (isomers of $C_{60}$),[60] presented insurmountable near-linear dependence problems (overlap matrix elements below $10^{-11}$) owing to the p diffuse functions and was eliminated: anyhow, it has a small weight in WTMAD2.

We performed PNO-DHDF-F12 calculations with the V$n$Z-F12 (where $n$ = D,T) basis sets. For six anion-containing subsets, AHB21,[61] G21EA,[62,63] IL16,[61] WATER27,[64,65] BH76,[6,63,66,67] and BH76RC,[6,63] as well as for the RG18[6] rare-gas clusters, we also considered aug-cc-pV$n$Z-F12 ($n$ = D,T)[33] or AV$n$Z-F12 for short; the combination of the latter with V$n$Z-F12 for the rest of the GMTKN55 suite is denoted V$n$Z-F12* as in ref [46].

For comparison, we also carried out conventional B2GP-PLYP calculations with the commonly used Weigend-Ahlrichs basis sets[68] def2-TZVPP and def2-QZVPP as well as with their diffuse-function augmented equivalents[69] def2-TZVPPD and def2-QZVPPD. Additionally, we considered basis sets of the correlation consistent family:[70−72] the shorthand haV$n$Z (heavy-augmented valence $n$-tuple zeta, where $n$ = D,T,Q,5) stands here for the combination of cc-pV$n$Z on hydrogen,[70] aug-cc-pV($n$+d)Z on second-row elements,[72] aug-cc-pV$n$Z on first-row[71] and third-row[73] elements, and aug-cc-pV$n$Z-PP on the fourth- and fifth-row p-block elements.[74−77] The shorthand V$n$Z stands for the variant without diffuse functions of this same combo.

The largest basis set for which we were previously able[46] to obtain fully canonical B2GP-PLYP-F12-D3(BJ) answers, permitting a direct comparison, was VQZ-F12*, permitting extrapolation from VTZ-F12* and VQZ-F12* or for short V{T,Q}Z-F12*.

The top section of Table 1 presents the WTMAD2 for localized PNO-B2GP-PLYP-F12-D3(BJ) for GMTKN55 and its breakdown into the five top-level categories. With the VDZ-F12 basis set, we obtained a WTMAD2 of 3.080 kcal/mol for the GMTKN55 data set, which goes down insignificantly to 3.073 kcal/mol for VDZ-F12*. Furthermore, VTZ-F12 yields a WTMAD2 of 3.030 kcal/mol, which goes down to 2.961 kcal/mol for the VTZ-F12* variant. We extrapolated VDZ-F12* and VTZ-F12* reaction energies using the two-point extrapolation formula ($A + B/L^\alpha$; $L$ = highest angular momentum present in the basis set) where $\alpha$ = 3.088 for the PT2 components (Table 9 in ref [78], first row of lower pane, second column); for the KS component, we just used the highest angular momentum present in the basis set plus the CABS correction. The PNO-B2GP-PLYP-F12-D3(BJ)/V-{D,T}Z-F12* level of theory results in WTMAD2 of 2.951 kcal/mol for the entire GMTKN55 database. The WTMAD2 results obtained with canonical B2GP-PLYP-F12-D3(BJ) are 2.939, 2.969, and 2.993 kcal/mol, respectively, for VDZ-F12*, VTZ-F12*, and V{D,T}Z-F12* (see ref [46]).

Next (Table 1, middle section), we explored the basis set convergence of localized B2GP-PLYP-F12-D3(BJ) relative to energies calculated at the canonical B2GP-PLYP-F12-D3(BJ)/V{T,Q}Z-F12* level of theory, which can essentially be regarded as the complete basis set limit. Relative to that, PNO-B2GP-PLYP-F12-D3(BJ) calculations in conjunction with VDZ-F12 and VDZ-F12* provide WTMAD2$_{CBS}$ values of 0.642 and 0.633 kcal/mol, respectively. Increasing the basis set size to VTZ-F12 and VTZ-F12* reduces these deviations to 0.322 and 0.233 kcal/mol, respectively. PNO-B2GP-PLYP-F12-D3(BJ) in conjunction with V{D,T}Z-F12 yields a WTMAD2$_{CBS}$ value of 0.300 kcal/mol. Adding diffuse functions to RG18, WATER27, BH76, BH76RC, AHB21, G21EA, and IL16 (i.e., V{D,T}Z-F12*) slightly lowers the WTMAD2$_{CBS}$ value to 0.262 kcal/mol. WTMAD2$_{CBS}$ values obtained for canonical B2GP-PLYP-F12-D3(BJ) with VDZ-F12*, VTZ-F12*, and V{D,T}Z-F12* basis sets are 0.467, 0.207, and 0.215 kcal/mol, respectively (see ref [46]).

Another angle is offered by considering the WTMAD2 between PNO-B2GP-PLYP-F12-D3(BJ) and canonical B2GP-PLYP-F12-D3(BJ) energies in the same basis set. These can be found in the bottom section of Table 1. For VDZ-F12, this "ΔWTMAD2$_{PNO}$" is 0.513 kcal/mol, most of which is accounted for by the intermolecular and conformational subsets (0.258 and 0.113 kcal/mol, respectively). RG18 alone contributes 0.154 kcal/mol, which together with 0.037 (HEAVY28), 0.024 (S66), and 0.018 (HAL59) accounts for nearly the entire INTERMOL component; CONFOR is mostly due to the four subsets PCONF21 (0.026), BUT14DIOL (0.025), Amino20x4 (0.024), and MCONF (0.019 kcal/mol). Essentially the same picture emerges for VDZ-F12*. For VTZ-F12, the "PNO ΔWTMAD2" is 0.333 kcal/mol, the largest two contributor sets being RG18 and HEAVY28 at 0.094 and 0.038 kcal/mol, respectively; in the VTZ-F12* variant, the RG18 contribution drops to 0.029 kcal/mol. A similar accuracy is found for V{D,T}Z-F12 and V{D,T}Z-F12* extrapolations; in each case, RG18 is again the largest contributor.

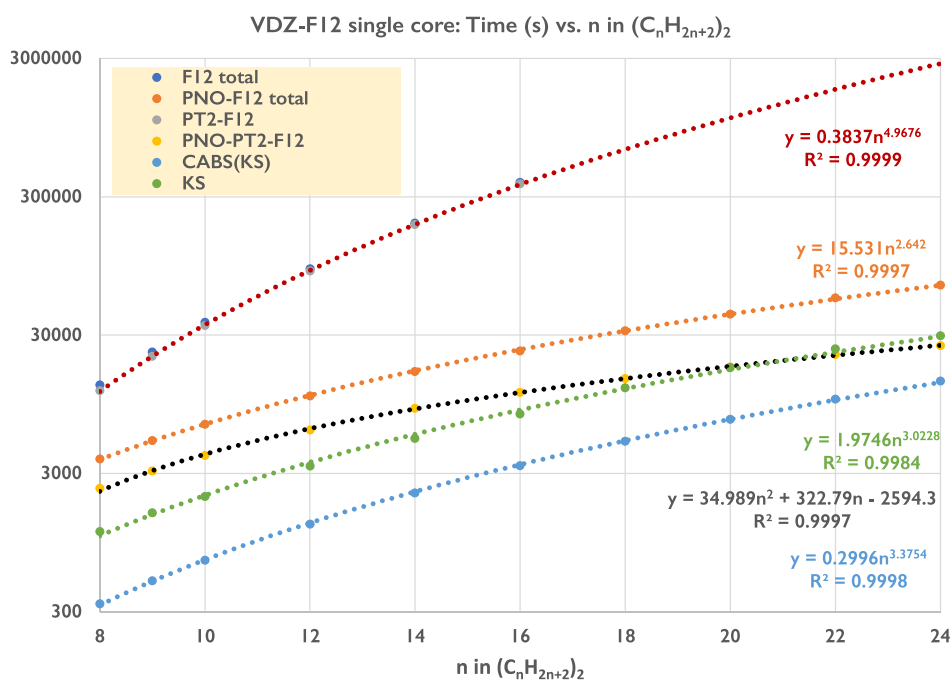

**Figure 1.** Elapsed times (s, logarithmic scale) of canonical GLPT2-F12 and localized PNO-GLPT2-F12 steps of $(C_nH_{2n+2})_2$ on a single Intel Haswell E5-2630 v3 core at 2.4 GHz with 256 GB RAM and 3.6TB striped SSD.

We shall now briefly compare the computational cost of localized (denoted PNO-F12 for short in the tables) and canonical B2GP-PLYP-F12-D3(BJ) calculations. This is perhaps best illustrated by considering timings for the linear n-alkane dimers as a function of n; we considered first (ethane)$_2$ through (n-heptane)$_2$, which make up the ADIM6[6,79] subset of GMTKN55. Then, we extended this series for $n = 8$ through $n = 24$ by manually inserting CH$_2$ groups (since we were only interested in timings, optimized structures were not deemed necessary). All calculations in this comparison are run on identical hardware, namely, Intel(R) Haswell 2.40 GHz, 16 cores, 256 GB RAM, and 3.6TB of fast striped SSDs. Detailed single-core and 16-core timings for both F12 and orbital-only calculations with various basis sets can be found in the Supporting Information; in Table 2, we report values relative to PNO-F12/VDZ-F12 = 1.00. As we have seen above, localized PNO-B2GP-PLYP-F12/VDZ-F12 yields results of comparable quality to canonical B2GP-PLYP-F12/VDZ-F12. The localized PNO-B2GP-PLYP-F12/VDZ-F12 results are comparable in cost to their canonical counterparts for n-butane dimer, but as the chain length grows, the gap opens up to the point where for (n-hexadecane)$_2$ the PNO calculation is almost 18 times faster than its canonical counterpart. The cost for PNO-B2GP-PLYP-F12/VTZ-F12 is an almost constant factor of 3.5−3.6 greater than for the same calculation in the smaller VDZ-F12 basis set—meaning that, for the n-decane dimer (the largest case for which we were able to do the canonical F12/VTZ-F12 calculation), the latter took about 6 times as long as its PNO counterpart. When compared with B2GP-PLYP/haV{T,Q}Z, PNO-B2GP-PLYP-F12/VDZ-F12 goes from about 2 times to about 7 times faster as the chains grow longer and, for def2-{T,Q}ZVPPD, from 2 to 5 times faster. The B2GP-PLYP/haV{Q,5}Z calculations range from about 9 times longer than PNO-B2GP-PLYP-F12/VDZ-F12 for the butane dimer to about 27 times as long for (n-hexadecane)$_2$.

In fact, a power law fit to the total computational time for n-heptane through n-hexadecane dimers, $(C_nH_{n+2})_2$ ($n = 7−10$, 12, 14, 16), reveals $\propto n^{4.88}$ scaling (nearly the expected $\propto n^5$) for canonical B2GP-PLYP-F12-D3(BJ) but approximately $\propto n^{2.58}$ scaling for PNO-B2GP-PLYP-F12-D3(BJ).

For mass storage requirements, the ratios are even more lopsided (see Table S1): the PNO-F12 calculation on the n-butane dimer requires only one-sixth the scratch space of its canonical counterpart, and for n-hexadecane, this drops to one-16th. These ratios appear pretty much independent of the basis set. What this means in concrete terms: for the n-hexadecane dimer with the VDZ-F12 basis set, the canonical calculation requires almost 3.5 TB of scratch space versus about 0.2 TB for the PNO calculation. On our cluster, this makes the difference between having to run the calculation on a dedicated "heavyio" node with a large local scratch SSD and being able to run it on any available general-purpose node.

Note that, for the conventional B2GP-PLYP-F12-D3(BJ)/ cc-pVDZ-F12 calculation, the PT2-F12 step dominates the CPU time to such an extent that the "F12 total" line is obscured by the crimson "PT2-F12" line. KS is the time spent in the Kohn−Sham iterations; CABS(KS) denotes that for the evaluation of the CABS correction. PT2-F12 and PNO-PT2-F12 refers to the canonical and PNO perturbation theory steps, respectively; "F12 total" is the time for the entire canonical calculation, and "PNO-F12 total" refers to the entire localized calculation.

However, for larger systems, CPU time scaling of the PNO-GLPT2-F12 step in PNO-B2GP-PLYP-F12-D3(BJ) becomes even gentler. Figure 1 illustrates this for dimers of parallel chains of n-alkanes through $n = 24$ (tetraicosane dimer) with the cc-pVDZ-F12 and cc-pVTZ-F12 basis sets. The canonical calculations were only feasible through $n = 16$ (for which the PT2-F12 step required all available scratch space): a power law fit of CPU times for the PT2-F12 step reveals an almost perfect $\propto n^5$ dependence ($R^2 = 0.9999$) as will the total CPU time that is completely dominated by this step. For PNO-B2GP-PLYP-F12-D3(BJ)/cc-pVDZ-F12, a power law fit for $n = 8$ through $n = 24$ reveals a much gentler $\propto n^{2.64}$ scaling ($R^2 = 0.9997$): a component breakdown of the times reveals an approximately $\propto n^3$ scaling for the Kohn−Sham step ($R^2 = 0.998$) paired with a scaling for the PNO-PT2-F12 step that follows a roughly $\propto n^{2.16}$ power law ($R^2 = 0.997$) but hews more closely ($R^2 = 0.9997$) to a quadratic fit. Broadly speaking, the same trends are found for cc-pVTZ-F12 (see the Excel sheet in the Supporting Information).

When running in a more realistic fashion on 16 cores, we find speedups by about a factor of 8 for B2GP-PLYP-F12-D3(BJ) and of 12 for PNO-B2GP-PLYP-F12-D3(BJ). More detailed scrutiny reveals that, while in the former case, the scaling is determined by the PT2-F12 step (where I/O bandwidth limitations place a practical limit on parallelism), in the latter case PNO-PT2-F12 actually parallelizes close to ideally and the Kohn−Sham step is the parallelization efficiency-limiting factor. At the end of the day, parallelization additionally favors PNO-B2GP-PLYP-F12-D3(BJ) and makes it an even more attractive method.

Now, would we be able to realize similar gains from PNO-LMP2 in orbital-only (i.e., non-F12) calculations? We have considered this for haVTZ and haVQZ basis sets along the same $n$-alkane dimer series for $n = 8$−16, parallel on 16 cores; the timing data and their breakdown can again be found in the Supporting Information. For low $n$, the PNO approach is actually slightly costlier, but for $n = 16$, we can see a reduction in total CPU time by about 25%. Needless to say, this does not even come close to the order-of-magnitude or more that can be saved in PNO-F12 vs canonical F12 double hybrids. (We note that, in the orbital-only calculations, due to the larger basis sets, the KS step accounts for the lion's share of the total time, although this will eventually be reversed as chains grow still longer.)

The principal conclusion of this work is that double hybrid F12 calculations, which largely eliminate the slow basis set convergence of double hybrids, can be carried out without significant loss of accuracy using localized pair natural orbitals in the F12 step. Thus, CPU and mass storage requirements scale much more gently with the system size, making the method amenable also to, and promising for, larger systems.

## ASSOCIATED CONTENT

### Supporting Information

The Supporting Information is available free of charge at https://pubs.acs.org/doi/10.1021/acs.jpclett.2c02620.

A Microsoft Excel workbook containing statistical results of all assessed methods as well as a sample MOLPRO input file for PNO-B2GP-PLYP-F12 and 1-core and 16-core timing data for the $n$-alkane dimer series using different approaches and broken down by calculation step (XLSX)

## AUTHOR INFORMATION

### Corresponding Author

Jan M. L. Martin − *Department of Molecular Chemistry and Materials Science, Weizmann Institute of Science, Reḥovot 7610001, Israel;* orcid.org/0000-0002-0005-5074; Phone: +972-8-9342533; Email: gershom@weizmann.ac.il; Fax: +972-8-9343029

### Author

Nisha Mehta − *Department of Molecular Chemistry and Materials Science, Weizmann Institute of Science, Reḥovot 7610001, Israel;* orcid.org/0000-0001-7222-4108

Complete contact information is available at:
https://pubs.acs.org/10.1021/acs.jpclett.2c02620

### Notes

The authors declare no competing financial interest.

## ACKNOWLEDGMENTS

Work on this paper was supported by the Israel Science Foundation (grant 1969/20), by the Minerva Foundation (grant 2020/05), and by a research grant from the Artificial Intelligence and Smart Materials Research Fund (in memory of Dr. Uriel Arnon), Israel. The authors would like to thank Golokesh Santra for helpful comments on the draft manuscript and Emmanouil Semidalas for proofreading the revised manuscript.